\documentclass[12pt,preprint]{aastex}

\usepackage{amsmath,amssymb}
\usepackage{natbib}
\usepackage{epsfig,graphics}
\usepackage{epstopdf} 

\shorttitle{First Detection of Near-Infrared Line Emission from Organics in Young Circumstellar Disks}
\shortauthors{Mandell, Bast, van Dishoeck, Blake, Salyk, Mumma, Villanueva}

\begin{document}
\title{First Detection of Near-Infrared Line Emission from Organics in Young Circumstellar Disks} 
\author{Avi M. Mandell\altaffilmark{1,7}, Jeanette Bast\altaffilmark{2}, 
Ewine F. van Dishoeck \altaffilmark{2,3}, Geoffrey A. Blake\altaffilmark{4}, Colette Salyk\altaffilmark{5},
Michael J. Mumma\altaffilmark{1}, Geronimo Villanueva\altaffilmark{1,6}} 
\altaffiltext{1}{Solar System Exploration Division, NASA Goddard Space Flight Center, Greenbelt, MD 20771, USA}
\altaffiltext{2}{Leiden Observatory, Leiden University, P.O. Box 9513, 2300 RA Leiden, The Netherlands}
\altaffiltext{3}{Max Planck Institut f\"ur extraterrestrische Physik, P.O. Box 1312, D-85741 Garching, Germany}
\altaffiltext{4}{California Institute of Technology, Division of Geological and Planetary Sciences, MS 150-21, Pasadena, CA 91125, USA}
\altaffiltext{5}{University of Texas, Department of Astronomy, Austin, TX 78712, USA}
\altaffiltext{6}{Catholic University of America, Department of Physics, Washington, DC 20064, USA}
\altaffiltext{7}{Corresponding Email:  Avi.Mandell@nasa.gov}

\begin{abstract}
We present an analysis of high-resolution spectroscopy of several bright T Tauri stars using the VLT/CRIRES and Keck/NIRSPEC spectrographs, revealing the first detections of emission from HCN and C$_2$H$_2$ in circumstellar disks at near-infrared wavelengths.  Using advanced data reduction techniques we achieve a dynamic range with respect to the disk continuum of $\sim500$ at 3\,$\mu$m, revealing multiple emission features of H$_2$O, OH, HCN, and C$_2$H$_2$.  We also present stringent upper limits for two other molecules thought to be abundant in the inner disk, CH$_4$ and NH$_3$. Line profiles for the different detected molecules are broad but centrally peaked in most cases, even for disks with previously determined inclinations of greater than 20$\degr$, suggesting that the emission has both a Keplerian and non-Keplerian component as observed previously for CO emission.  We apply two different modeling strategies to constrain the molecular abundances and temperatures: we use a simplified single-temperature LTE slab model with a Gaussian line profile to make line identifications and determine a best-fit temperature and initial abundance ratios, and we compare these values with constraints derived from a detailed disk radiative transfer model assuming LTE excitation but utilizing a realistic temperature and density structure.  Abundance ratios from both sets of models are consistent with each other and consistent with expected values from theoretical chemical models, and analysis of the line shapes suggests the molecular emission originates from within a narrow region in the inner disk (R $<1$ AU).  \footnote{Based partially on observations collected at the European Southern Observatory Very Large Telescope under program ID 179.C-0151, program ID 283.C-5016, and program ID 082.C-0432 (PI: Pontoppidan).}
\end{abstract}


\section{Introduction}
\label{intro} 
The chemical and thermal distribution of the warm gas in the inner regions of protoplanetary disks provides important diagnostics related to the composition of the material that formed into planets around nearby stars. With the recent discovery of surprisingly large numbers of super-Earths (planets with masses of $1-10$ M$_{Earth}$) within a few AU of their parent stars, significant questions have been raised concerning their origin and composition \citep{Borucki2011p19}. Spectroscopy of the atmospheres of a few such super-Earth and mini-Neptune planets show large variations in composition from object to object \citep{Madhusudhan2011p41,   Desert2011pL40}, and it is clearly important to constrain the chemistry of the gas in the inner regions of protoplanetary disks in order to better understand the origin of the atmospheres and interiors of these planets.

With improved infrared instrumentation on ground- and space-based telescopes, various molecules can now be observed in the warm irradiated surface layers of disks, allowing us to determine their chemistry and dynamics. CO rovibrational emission at 4.7\,$\mu$m has been extensively surveyed in both Herbig Ae (HAe) and T Tauri stars \citep{Brittain2003p535, Najita2003p931, Blake2004pL73, Pontoppidan2008p1323, Salyk2009p330, pontoppidan2011p84, Bast2011p119}, H$_2$O emission has been detected in both the overtone \citep{Carr2004p213} and fundamental bands \citep{Salyk2008pL49}, and there have been early detections of additional tracers such as warm H$_2$ \citep{Bitner2007pL69} and OH \citep{mandell2008pL25, Fedele2011p106}. Just in the last few years, there have been surprising detections of absorption by C$_2$H$_2$, HCN and CO$_2$ in systems with unique viewing geometries \citep{Lahuis2006pL145, Gibb2007p1572}, and high S/N data from {\it Spitzer} are now revealing an extraordinarily rich mid-IR spectrum in emission containing lines of water, OH and organics \citep{Carr2008p1504, Carr2011p102, Pascucci2009p143, Pontoppidan2010p887, Salyk2011p130}. Overall, we are moving toward a comprehensive picture of the disk chemistry and temperature structure in the region of the disk in which planets form; these results will aid in understanding processes such as planetary migration, disk turbulence and the accretion of volatiles. However, to accurately constrain the physical structure and chemical balance among different molecular species in different regions of the disk we must analyze a full suite of molecular tracers, with transitions excited at both high and low temperatures, using instruments with sufficient resolving power to constrain different line shapes indicative of different Keplerian velocities (i.e., different source regions).

In this paper we present high-resolution ($R=25,000 - 95,000$) L-band ($3-4\,\mu$m) observations of three bright T Tauri stars using the CRIRES instrument on the Very Large Telescope and the NIRSPEC instrument on the Keck II telescope, with the goal of further characterizing molecular emission from gas tracers that have been detected with the IRS instrument on {\it Spitzer} (primarily HCN) and to search for new species not covered by the {\it Spitzer} bandpass, specifically CH$_4$. The latter two molecules are observed to be significant components of interstellar ices \citep{Oberg2008p1032, Bottinelli2010p1100} and of comets \citep{Mumma2011p471}, with typical abundances of a few percent with respect to H$_2$O. If incorporated unaltered into icy planetesimals in the cold outer regions and subsequently transported inward across the snow line, they become part of the gas in the planet-forming zones. Observations of CH$_4$ are of particular interest because of the recent controversies on the CH$_4$/CO abundance ratios in exoplanetary atmospheres \citep{Madhusudhan2011p64, Madhusudhan2011p41}.

The search for new molecules is best done toward sources which show high fluxes of common molecules and high line-to-continuum ratios. The three stars observed for this study have all been identified through their unusually bright CO emission \citep{Bast2011p119}, and two of them (AS 205 A and DR Tau) have been shown to have bright water emission at 3.0\,$\mu$m \citep{Salyk2008pL49}. AS 205 A is a K5 star within a triple system in the constellation of Ophiuchus.  With an age of 0.5 Myr and a disk inner radius of 0.07 to 0.14 AU, the system is the archetype of a young gas-rich accreting primordial disk \citep{Andrews2009p1502,Eisner2005p952}.  DR Tau is a K7 star, while RU Lup is a G5 star; these disks are less bright at NIR wavelengths than AS 205 A but similar in their rich molecular emission characteristics.  We present a preliminary analysis of the detected molecular disk tracers using several different molecular emission models, discuss our results in the context of recent observations of mid-IR emission from the same objects as well as theoretical models of disk chemistry, and describe future work required to improve our constraints on the abundance and distribution of these molecular gas tracers of the warm inner disk.

\section{Observations and Data Reduction}

\label{obs} Spectra were acquired over several observing epochs using the CRIRES instrument on the VLT ($\lambda/\Delta\lambda \approx $ 96,000) as part of our large program to study primordial circumstellar disks \citep{pontoppidan2011p32}, and additional data were acquired with the NIRSPEC instrument on Keck ($\lambda/\Delta\lambda \approx $ 25,000). Comparison stars were also observed as close in time and airmass to the science targets as possible; a full list of targets is listed in Table \ref{tab:obs}. Both telescopes were nodded in an ABBA sequence with a throw of $10-12$ arcseconds between nods, with a 60-second integration time per image (or combination of co-adds).  The CRIRES observations for DR Tau and RU Lup were taken using two different wavelength settings in order to cover the gaps between the instrument detectors, while only a single setting was used for AS 205 A; a single echelle and cross-disperser setting was used for the NIRSPEC observations of DR Tau.  In total, we analysed 48 CRIRES spectra and 28 NIRSPEC spectra of DR Tau, 32 CRIRES spectra of RU Lup, and 36 CRIRES spectra of AS 205 A.

We utilized custom data reduction algorithms, previously used to detect new molecular emission features from warm gas in circumstellar disks \citep{mandell2008pL25, mandell2011p18} to extract and process spectra for each echelle order in each ABBA set. We reduced the initial 2D spectral-spatial images to 1D spectra after first correcting for the slope of the beam due to cross-dispersion and subtracting A- and B-beam images to remove the contribution from telluric radiance. We identified bad pixels and cosmic ray hits in each raw pixel column by comparing the beam profile to an average beam profile for nearby columns, allowing us to identify and replace single-pixel events without removing any enhancements due to emission or absorption features.  

\begin{figure}[htb]
\centering
{
 \includegraphics[width=170mm]{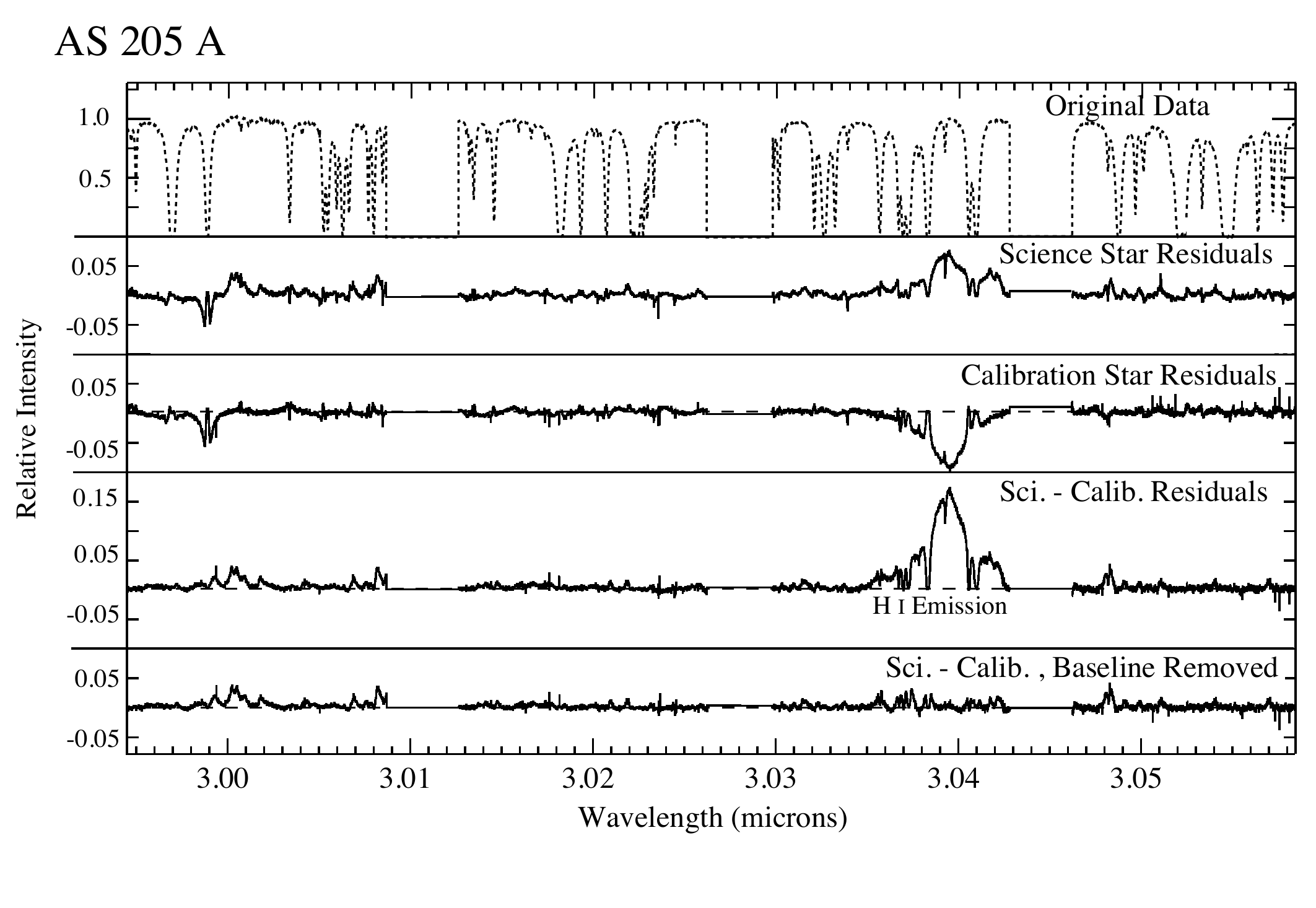}
 \caption{Data for AS 205 A, with the original data plotted in the first panel and subsequent residuals plotted below. The Science and Calibration Star Residuals are produced by removing a best-fit atmospheric model for each star separately; the two residuals are then differenced to produce the Science - Calibration Residuals.  A segmented baseline composed of several polynomial fits is then used to remove broad features such as the H{\scriptsize{ I}} emission line.}
 \label{fig:data_raw}
}
\end{figure}

We corrected for changing airmass and telluric atmospheric conditions by fitting the data with a terrestrial spectral transmittance model synthesized with the LBLRTM atmospheric code \citep{clough2005p233} based on line parameters from the HITRAN 2008 molecular database with updates from 2009 \citep{rothman2009p533}. The telluric models were optimized to fit the observational data by varying the abundances of the primary atmospheric constituents (H$_2$O, CO$_2$, CH$_4$, O$_3$, and N$_2$O), with the resolving power of the model set by the width of the narrowest telluric features. Atmospheric models were fitted for wavelength sub-sections of each AB set, and the telluric model was subtracted from the stellar data to produce a first set of residuals. We performed the same telluric removal routine on the comparison star, and the post-telluric residuals of the science and comparison stars were then differenced to remove remnant fringes and other instrumental artifacts, as well as minor errors in the telluric model such as improper pressure broadening and isotopic ratios. The residuals for all sections were then combined, and the complete spectrum was then fit with a new segmented continuum model to remove obvious broad emission features, most notably the H{\scriptsize{ I}} emission line at 3.04\,$\mu$m. The new continuum baseline matched a polynomial fit to individual segments of the spectrum whose length was defined interactively (but not less than 100 pixels), allowing us to fit sections of the spectrum with a complex morphology.  The different steps of the reduction process are illustrated in Figure~\ref{fig:data_raw}.

The reduction process yields spectra with S/N on the stellar continuum corresponding to an rms noise only slightly larger (by 20\%) than that expected from the photon statistics, reaching $\sim 500$ for AS 205 A. These are the deepest disk spectra obtained in this wavelength range with CRIRES to date. Data with remaining bad pixels and other artifacts (such as several unidentified stellar features in the comparison star for AS 205 A) were removed; the regions were 5 pixels or less, and did not affect the processing or analysis of nearby pixels.  Flux calibration was performed by normalizing the observed continuum flux to the predicted flux based on the K-band stellar magnitude \citep{cutri2003p?}, which was then converted to L-band magnitudes using colors from the literature \citep{Glass1974p237, Hartigan1990pL25, Appenzeller1983p291}.  The flux was further adjusted for a wavelength-specific flux assuming a continuum blackbody temperature of 1500~K, which is the standard temperature for the veiling continuum of T Tauri stars; however, since the stellar magnitudes were converted to L-band magnitudes using observed colors, this correction was relatively minor.

\section{Line Identification and LTE Slab Modeling}
\label{slab}
\begin{figure}[htb]
\centering
{
 \includegraphics[width=170mm]{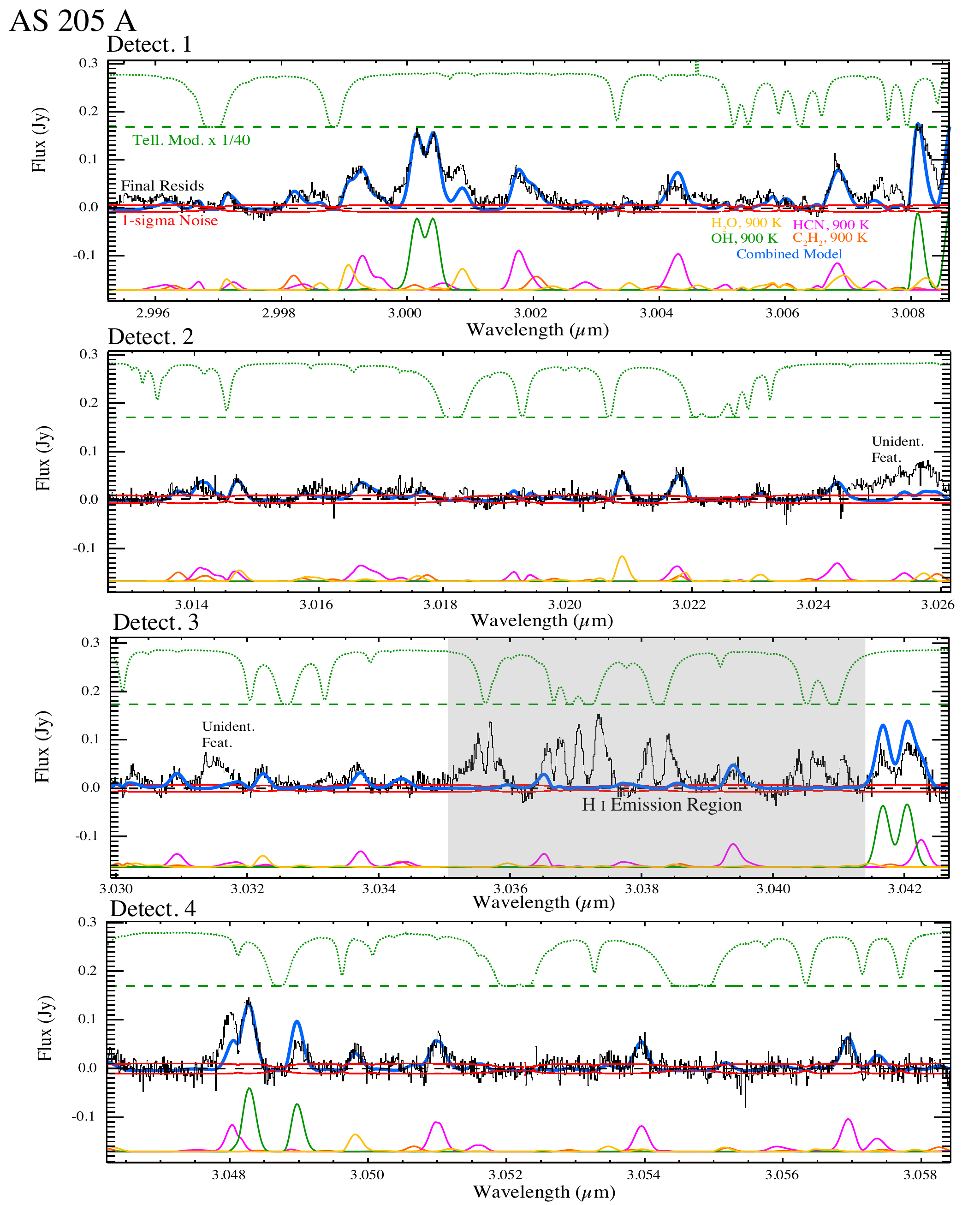}
 \caption{VLT-CRIRES data for AS 205 A overplotted with model spectra of LTE molecular emission from a simple slab model convolved with a Gaussian line shape. Emission from H$_2$O, OH, and HCN is clearly detected; emission features from C$_2$H$_2$ are also detected but the lines are weaker and blending with other features makes the detection less secure.}
 \label{fig:data_mols}
}
\end{figure}

\begin{figure}[htb]
\centering
{
 \includegraphics[width=170mm]{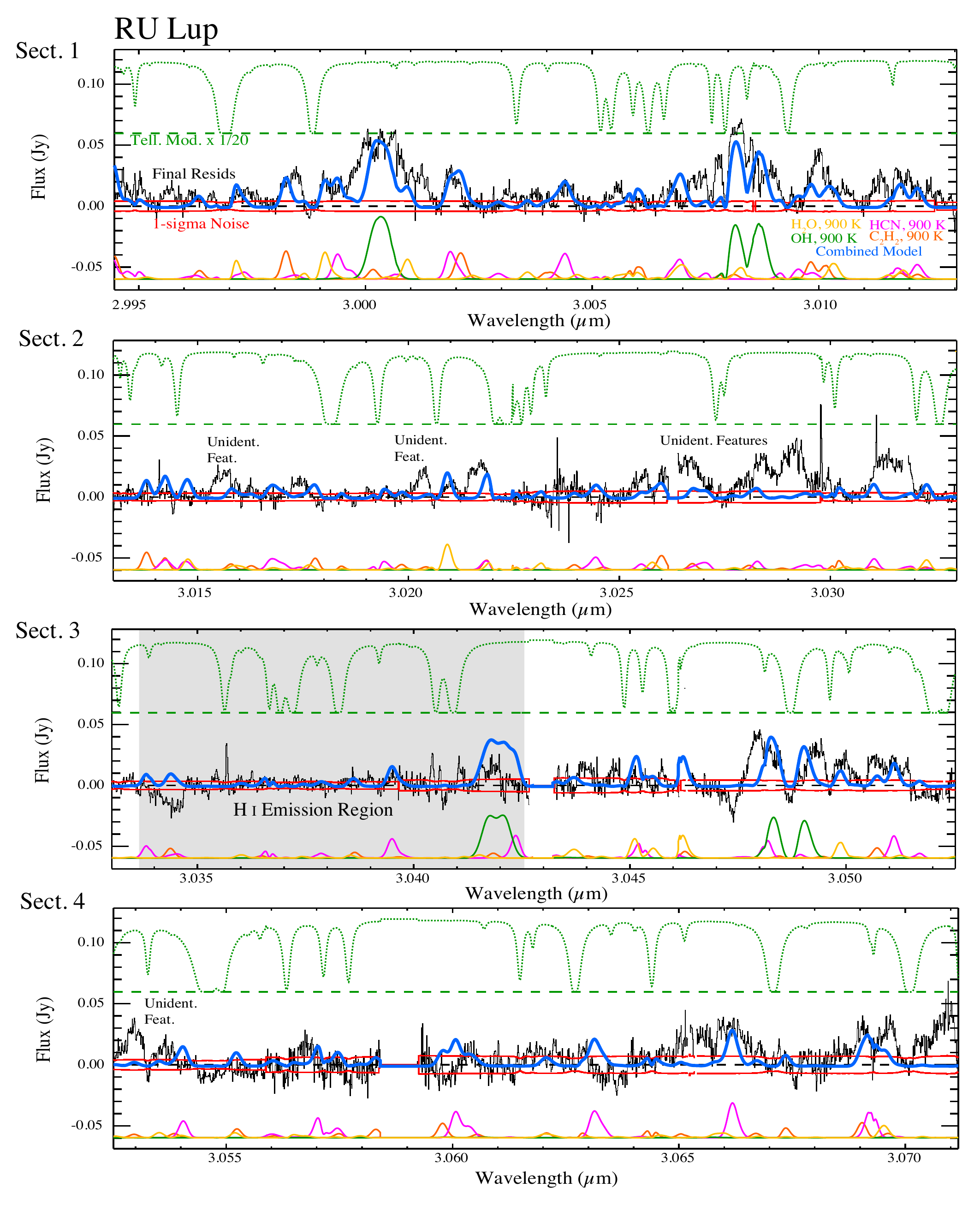}
 \caption{VLT-CRIRES data for RU Lup overplotted with model spectra from an LTE slab model. Emission from H$_2$O, OH, and HCN is present, but emission features from C$_2$H$_2$ are less secure than for AS 205 A.}
 \label{fig:data_mols_ru}
}
\end{figure}

\begin{figure}[htb]
\centering
{
 \includegraphics[width=170mm]{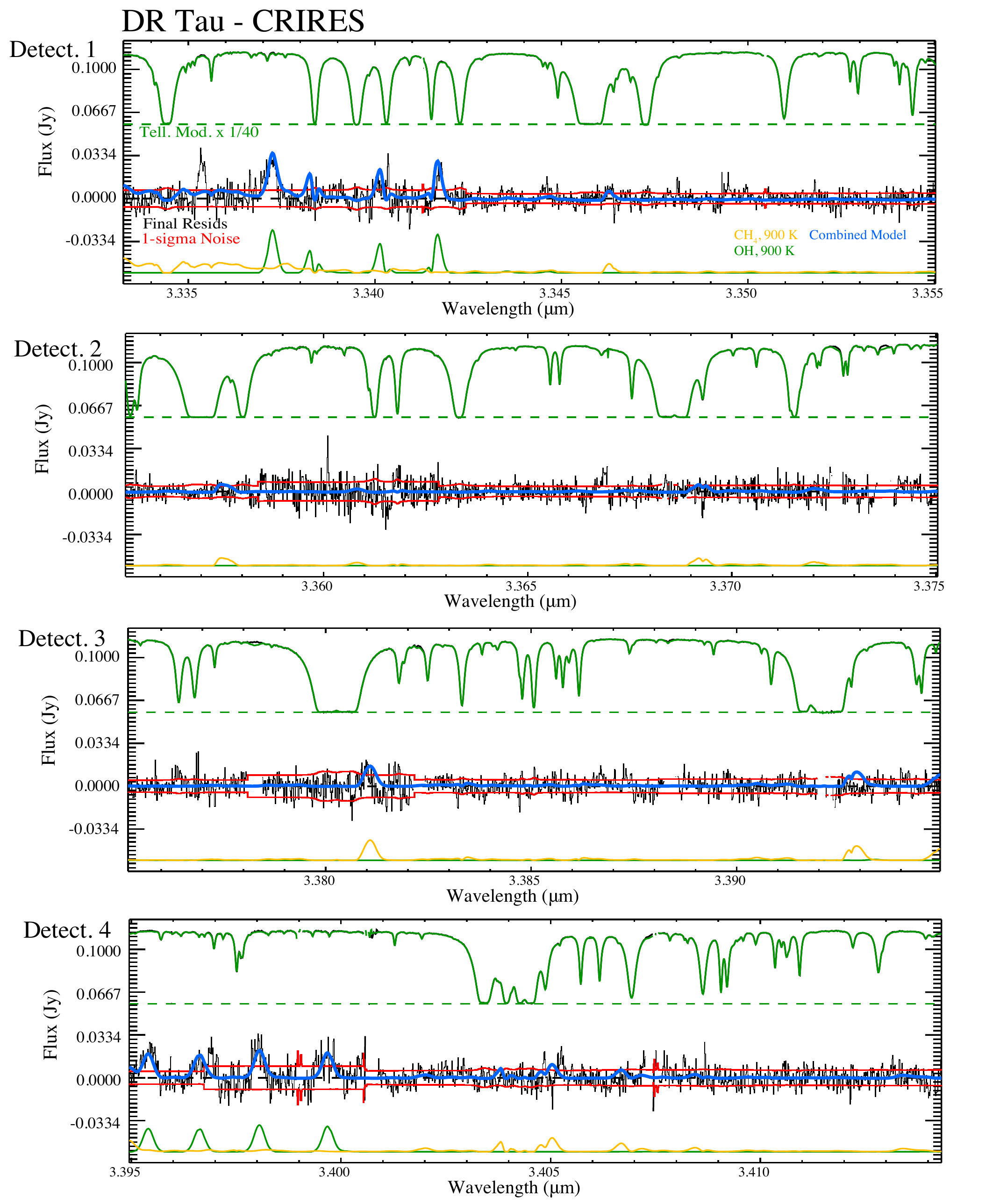}
 \caption{VLT-CRIRES data for DR Tau overplotted with model spectra for OH and CH$_4$ from an LTE slab model. Several high-J OH lines are clearly detected, but the evidence for CH$_4$ is not conclusive.  There is a possible features at 3.346 and 3.381\,$\mu$m, but other features at 3.358 and 3.393\,$\mu$m are not present.  We therefore only list an upper limit for the abundance of CH$_4$.}
 \label{fig:data_mols_dr}
}
\end{figure}

\begin{figure}[htb]
\centering
{
 \includegraphics[width=170mm]{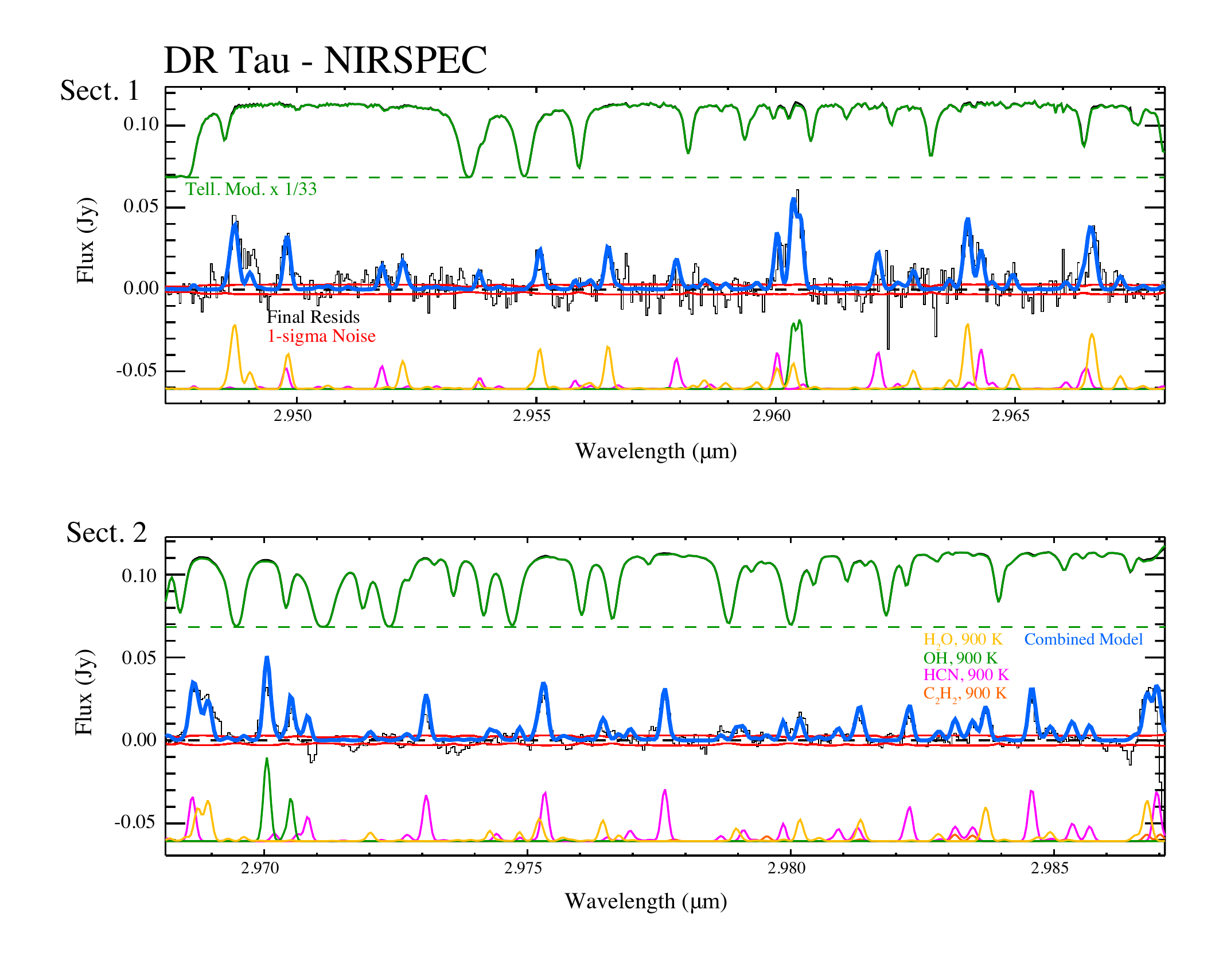}
 \caption{Keck-NIRSPEC Data for DR Tau overplotted with model spectra from an LTE slab model. Emission from H$_2$O, OH, and HCN is present, but emission features from C$_2$H$_2$ are too weak to be detected.}
 \label{fig:data_mols_dr_keck}
}
\end{figure}

The final spectral residuals for each star revealing disk emission are shown in Figures \ref{fig:data_mols} - \ref{fig:data_mols_dr_keck}.  The data for AS 205 A and RU Lup cover the spectral region between 2.995\,$\mu$m and 3.071\,$\mu$m, allowing us to search for multiple transitions of HCN, H$_2$O, OH, and C$_2$H$_2$, and emission from the $\nu_{1}$ band of NH$_3$. The NIRSPEC data for DR Tau cover the spectral region between 2.947\,$\mu$m and 2.987\,$\mu$m (spanning transitions of HCN, H$_2$O and OH), while the CRIRES data for DR Tau cover the region between 3.333\,$\mu$m and 3.414\,$\mu$m, allowing us to search for emission from OH and the $P$ branch of the 2$\nu_{2}$ band of CH$_4$ (Table \ref{tab:obs}).

\begin{figure}[htb]
\centering
{
 \includegraphics[width=170mm]{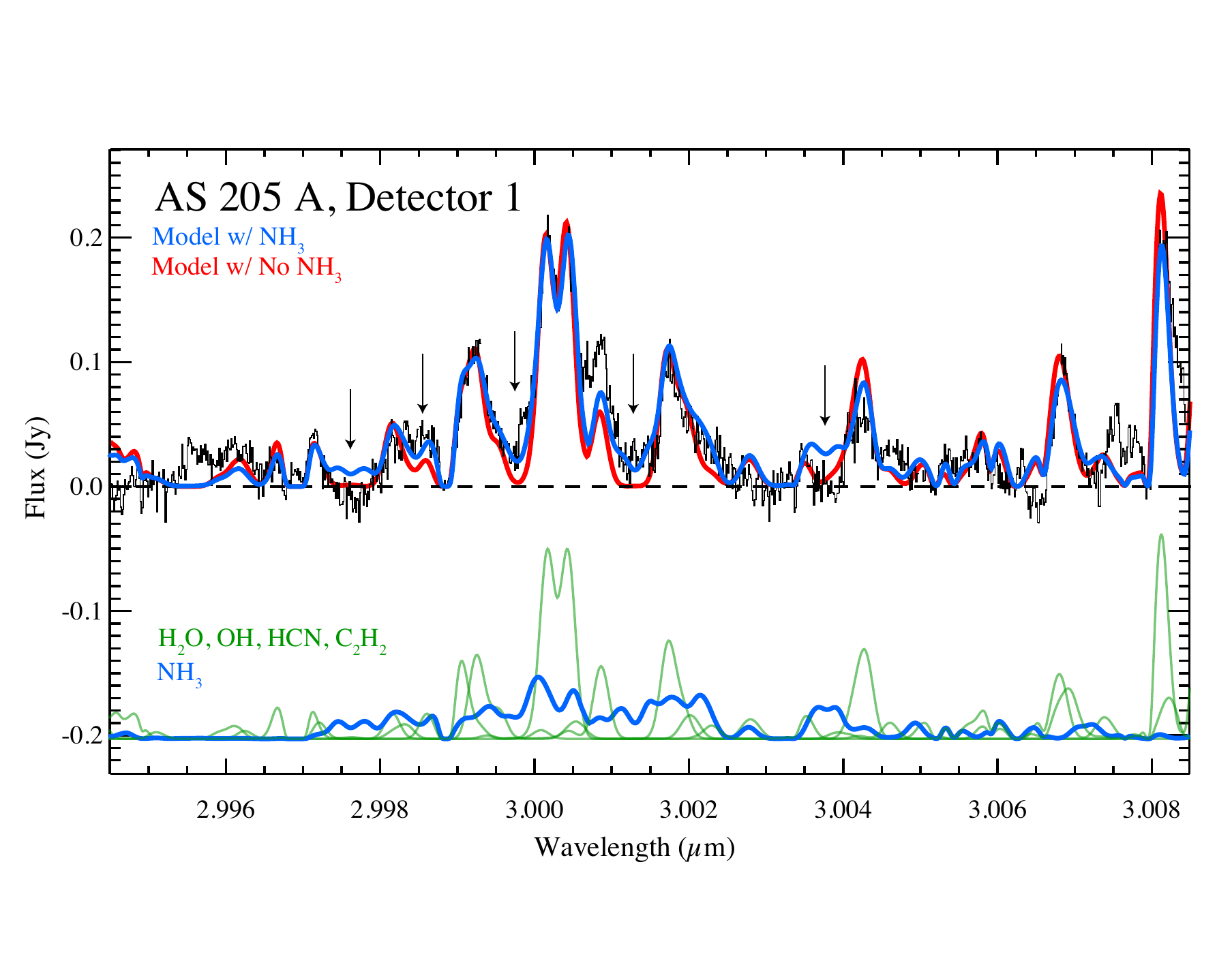}
 \caption{VLT-CRIRES data for AS 205 A, with best-fit models plotted both with and without the contribution from NH$_3$. The arrows identify
 locations where the models diverge, but the evidence is inconclusive as to whether emission from NH$_3$ is detected;
 we therefore only quote an upper limit.}
 \label{fig:nh3}
}
\end{figure}

Initial identifications of emission features in the residual spectra were assigned using data from the HITRAN 2008 molecular database with updates from 2009 \citep{rothman2009p533}.  We conclusively detect emission from H$_2$O, OH, and HCN in the spectra of all three stars, with multiple strong features of each molecule detected.  We also detect several low-amplitude features due to C$_2$H$_2$ in AS 205 A; there is one clean feature at 3.0137\,$\mu$m and several blended features (2.9981\,$\mu$m, 3.0021\,$\mu$m, 3.0177\,$\mu$m) that require C$_2$H$_2$ for a decent fit.  Part of the blend at 2.9981\,$\mu$m in the RU Lup spectrum may also be due to C$_2$H$_2$, but no other features are detectable. We can place an upper limit on CH$_4$ emission from DR Tau based on the CRIRES data; even though a weak feature is visible at 3.381\,$\mu$m, no corresponding lines are present at 3.393\,$\mu$m and 3.405\,$\mu$m that could be ascribed convincingly to CH$_4$. We also place upper limits on the presence of NH$_3$ emission from AS 205 A and RU Lup. The contribution of NH$_3$ at 2.9995\,$\mu$m and 3.0015\,$\mu$m improves the model fit, but there are several other locations where the fit is degraded (see Figure~\ref{fig:nh3}).  A list of selected lines for each molecule and the derived intensities for each star are listed in Table \ref{tab:lines}.

Initial abundance ratios for the molecules sampled in each spectrum were determined using a simple model of molecular gas in local thermal equilibrium (LTE), which approximates an optically thin medium where the line intensities are defined purely by the temperature of the emitting gas (i.e., a Boltzmann distribution) and the relative probability of each transition; this methodology is commonly known as a ``slab'' model.  We calculated the best-fit common temperature for the detected molecular features by varying the characteristic temperature between 300 K and 2000 K with a step of 50 K; we chose to use a common temperature in these fits due to the extensive blending of features and the narrow wavelength range being investigated. A temperature of 900 K was found to fit the relative line strengths for all species to within the uncertainties related to the complexity of the spectrum. Abundances for each molecule were then varied to fit the equivalent widths of the strongest lines.  The ratio of the abundance of each molecule relative to H$_2$O is listed in Table \ref{tab:ratios}.  It can be seen from Figures \ref{fig:data_mols} - \ref{fig:data_mols_dr_keck} that the basic LTE models do quite well in modeling the line positions and intensities, and very few features in the observed spectrum were left unidentified. Several residual features close to 3.035\,$\mu$m in both AS 205 A and RU Lup are due to the difficulty in fitting the broad H{\scriptsize{ I}} emission feature seen in Figure~\ref{fig:data_raw}. The feature at 3.0315\,$\mu$m in the spectrum of AS 205 A has no identification, and the features between 3.025 and 3.032 micron in both RU Lup and AS 205 A are also unidentified. 

A list of molecular ratios and upper limits referenced to H$_2$O are listed in Table \ref{tab:ratios}. We choose H$_2$O as the reference for the abundances because the features for all three molecules occur in the same wavelength region and the similar line shapes suggest they trace approximately the same region of the disk. We also include the ratio of CO to H$_2$O, calculated by fitting our LTE slab model to previously reduced CO spectra from CRIRES reported in \citet{Bast2011p119}. Our slab modeling is very similar to the methodology used in \citet{Bast2011p119}, and we refer the reader to that paper for figures and additional details.  We derived our CO abundance by fitting lines of $^{13}$CO and then multiplied the value by 65 \citep{Langer1990p477} to derive the overall CO abundance; the $^{13}$CO were explicitly shown to be optically thin for DR Tau \citet{Bast2011p119}, and we maintain the same assumption for AS 205 A and RU Lup (the CO data for these sources does not allow for confirmation of the optical depth, as discussed in \citet{Bast2011p119}).  We use a common rotational temperature of 600 K (close to the values found for AS 205 A and DR Tau by ) for consistency with the common temperature derived for the other molecules.  The difference in the best-fit temperatures for CO compared with the other molecules is robust and may be related to differences in emitting location (see \S\ref{result_model}), but a detailed analysis of the molecular distributions is reserved for future work.

\begin{figure}[htb]
\centering
{
 \includegraphics[width=110mm]{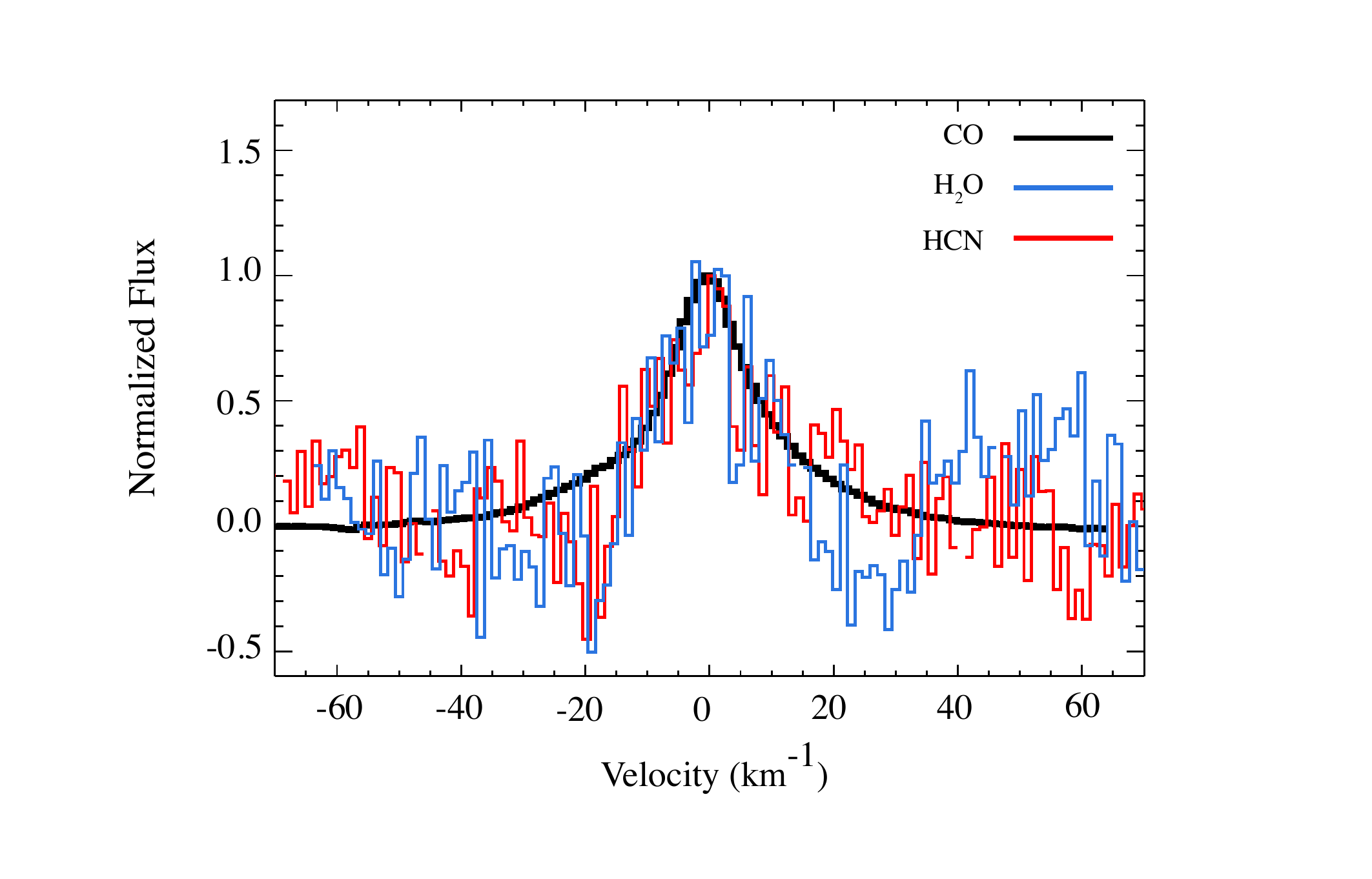}
 \caption{Rovibrational emission lines from three different molecules observed in AS 205 A. The different $J$-transitions are: P(8) (CO), P(11) (HCN) and (100)8$_{45}$ - (000)9$_{54}$ (H$_2$O). All three lines show the same shape to within measurement uncertainties, suggesting a common emitting distribution in the disk and/or the disk wind.}
 \label{fig:line_data}
}
\end{figure}

\begin{figure}[htb]
\centering
{
 \includegraphics[width=170mm]{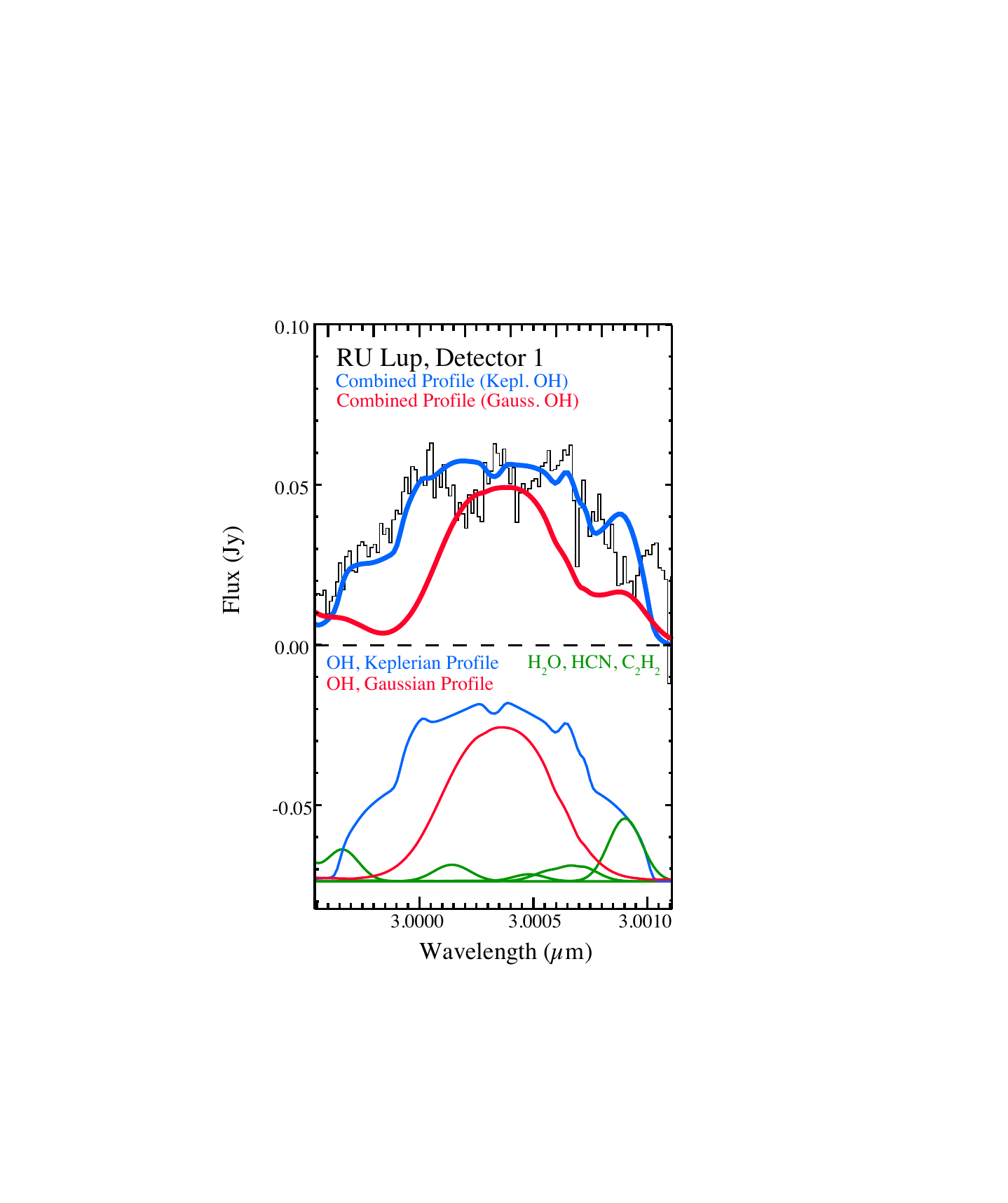}
 \caption{Profile for the OH transitions at 3.0\,$\mu$m for RU Lup, with two different models for the line shape overplotted. 
 A Keplerian profile with a steep decline in surface density and an outer radius of 0.5 AU clearly fits better than the simple Gaussian profile.}
 \label{fig:ohline}
}
\end{figure}

Previously determined inclinations for these objects from spectroastrometry range from $10\degr$ to $35\degr$ (\citealt{pontoppidan2011p84}; see Table \ref{tab:parameters1}), and the spectra therefore should show Keplerian line shapes if the outer radius of the emitting gas is restricted to the inner disk.  However, the line shapes for the newly detected molecules can be approximated quite well with a Gaussian profile, similar to the cores of the CO lines fit by \citet{Bast2011p119}. Figure~\ref{fig:line_data} shows single lines of HCN, H$_2$O and CO for AS 205 A normalized and plotted on top of each other. The HCN and H$_2$O lines are noisier due to their lower S/N but it can clearly be seen that the central regions of the three lines all have a similar single peaked line shape (we cannot stack features for improved S/N due to the extensive blending of features). Assuming a Gaussian line shape including thermal broadening and convolved with the instrumental profile, we find best-fit FWHM values between 10 km s$^{-1}$ and 22.5 km s$^{-1}$ for the strongest H$_2$O and HCN transitions in the three stars (see Table \ref{tab:ratios}).  The one exception to the single-peaked profile is the OH emission from RU Lup, whose features are clearly not fit well by a Gaussian model.  Figure~\ref{fig:ohline} demonstrates that a simple disk profile using an inner and outer radius ($0.05-0.5$ AU) and an exponential drop-off in flux ($I \propto R^{-3}$) provides a much better fit than the best-fit Gaussian profile for this OH line, suggesting a distribution limited to the very inner regions of the disk.  This corresponds well with the limit on the outer radius of 0.5 AU for OH and H$_2$O emission from AS 205 A determined by \citet{pontoppidan2011p84}, and a similar result was found by \citet{Fedele2011p106} for the OH lines from at least one HAe star.

As mentioned above, in the context of a Keplerian model the lack of a resolved double peak within the central core of most of the emission features in the CRIRES data would suggest significant emission from the outer disk (especially for disks with higher inclination such as RU Lup), and with the high spatial resolution of the VLT this emission would be detectable as an enhanced PSF size within the molecular line core compared with the continuum region outside the line. The line-to-continuum ratio for the H$_2$O and HCN features in our own spectrum is too low to allow a measurement of any deviations in the spatial profile due to extended emission, but both \citet{Bast2011p119} and \citet{pontoppidan2011p84} demonstrate that the measurements of the CO lines show no such emission.  These authors also show that the CO line shapes can actually be optimally modeled using a combination of a broad, low-amplitude (double-peaked) profile and a stronger single-peaked line core, suggesting a two-component profile. Recent spectroastrometric modeling of the CO features suggests that emission from a slow disk wind may account for the sub-Keplerian velocity profile of the central core of the line \citet{pontoppidan2011p84}. The disk wind explanation for the bulk of our emission is further supported by the fact that the emission features in each star appear to be shifted from the predicted position based on the stellar radial velocity by $\sim5$ km s$^{-1}$ (red-shifted for AS 205 A, blue-shifted for RU Lup), matching the shifts calculated for the narrow component of the CO profile in \citet{Bast2011p119}; however, line blending and uncertainties in line positions make the determination of velocity shifts difficult to confirm. Since the line profile will be determined by a combination of emission from both the disk and the wind, the similarity in line profiles of all three molecules suggests that the abundance ratios do not vary much between the disk and wind. We discuss the quantitative application of the disk wind model to our data in more detail in the following sections. 

The HCN/H$_2$O, OH/H$_2$O and CO/H$_2$O ratios derived here are all approximately an order of magnitude higher than the values derived by Salyk et al. (2011; hereafter S11) for the same sources using mid-infrared spectra taken with IRS/{\em Spitzer}.  Our upper limits for NH$_3$ with respect to water are also at least an order of magnitude higher than the {\it Spitzer}-based upper limits of $\leq0.01$ reported by S11. However, the usefulness of a direct comparison between the values derived by both slab model-based analyses is limited by several factors.   First, it is important to note that emission from molecular gas originates from a range of radial and vertical positions in the disk, and that a different portion of this range is probed by each wavelength regime.  Thus, there is no {\em requirement} that the molecular abundance ratios remain the same at each wavelength observed, and we could, in theory, observe radial variation in disk chemistry.  Nevertheless, one obvious limitation of both studies may resolve much of the discrepancy - the assumption of a common emitting location for all molecules.  This may especially affect the mid-IR abundance ratios, in which the pure-rotational water lines appear to probe lower temperatures ($300-500$ K), and hence larger disk radii, than the other molecules observed in the {\em Spitzer} spectra ($700-1100$ K).  Indeed, analyses of IRS/{\em Spitzer} data by \citet{Carr2008p1504, Carr2011p102} find smaller emitting areas for all molecules relative to H$_2$O and derive abundance ratios closer to those found here.  Similarly, the IRS/{\em Spitzer} wavelength range covers transitions with much lower excitation energies than the transitions probed by NIR observations, and therefore the optical depths measured will be different. Even at the same radial location the {\em Spitzer} observations may probe a different vertical location of the disk than the NIR observations. The abundance ratios for the different wavelength regions are also affected differently by the contribution of non-LTE excitation, since radiative pumping and collisional excitation processes can affect different transitions of various molecules in different ways. The OH/H$_2$O ratio is particularly uncertain since some of the OH emission may be radiatively excited and/or due to direct production into an excited state following photodissociation of H$_2$O \citep{Bonev2006p788}; the H$_2$O pure rotational mid-infrared lines are likely to be dominated by collisional excitation whereas the near-infrared lines are likely to be dominated by radiative excitation.

The limitations of the slab model approach highlight the need for proper treatment of the disk geometry and line radiative transfer. In the next section, we address the first slab-model limitation by modeling the near-infrared emission with a disk model including line radiative transfer, in which the emission derives from a range of disk radii and depths.  Non-LTE excitation requires a much more sophisticated approach, and we leave this analysis for future work.

\section{Disk Radiative Transfer Modeling}
\label{diskmod}

Since disks are not single-temperature slabs, we have also investigated the molecular emission from a two-dimensional radiative transfer disk model that takes into account both radial and vertical temperature and density variations, the motion of the gas, and the geometry of the disk. The modeling procedure for producing the line intensities and profiles centers around the axisymmetric ray-tracing code called RADLite, the details of which are described extensively in \citet{pontoppidan2009p1482}. RADLite calculates the emission intensity and the spectral line profile for a specific molecular transition by combining the emission from a grid of points across the projected surface of the disk.  This allows us to accurately reproduce the effects of the Keplerian rotation of the disk and the radial surface density and temperature profiles of each molecular tracer.  The initial temperature and density structure for the disk is calculated using the two-dimensional Monte-Carlo radiative transfer code called RADMC \citep{Dullemond2004p159}, and the dust temperature and source functions for each grid point are generated and then used as input for RADLite. In addition RADMC also calculates the continuum of the source. 

RADLite and RADMC offer the opportunity to test a number of different parameters that may affect the disk structure and resulting spectral model, and we present results from models that incorporate several different molecular abundance distributions.  Our fiducial model (Model 1) assumes a constant radial and vertical abundance for every molecule, with disk parameters taken from previous observational constraints, primarily from spectral energy distribution fits and sub-millimeter observations. The exact values of the parameters used for the model for each source are presented in Table \ref{tab:parameters1}. The model parameters used in RADMC are the mass of the central source ($M_\star$), its radius ($R_\star$) and effective temperature ($T_{eff}$), the mass of the disk ($M_{disk}$), outer radius of the disk ($R_{out}$), a flaring parameter $H$/$R$ $\propto$ $R^{\alpha}$, an outer pressure scale height ($h_{p}$/$R$), radial surface density $\Sigma$ = $R^\beta$ and an inner radius ($R_i$) which sets the inner temperature. The thermal broadening is assumed to dominate the intrinsic line width with the turbulent velocity $v_{turb}$ set to $v_{turb}$ = $\epsilon$$c_s$ with $\epsilon$ = 0.03. The inclination of the disk $i$ is taken from spectro-astrometry of \citet{pontoppidan2011p84} and is slightly adjusted (by less than 10\,$^{\circ}$) so a good fit can be found between the model and the data.  Model 1 also assumes an effective gas-to-dust ratio of 12800, based on previous modeling of water emission for T Tauri stars using RADLite \citep{Meijerink2009p1471}. This is much higher than the canonical value of 100, but can be explained by the depletion of small dust grains from the surface layers of the disk where our detected emission originates from.

In addition to our fiducial disk model, we varied the abundance structure in RADMC in three different ways in order to test the effects of freeze-out and dust settling.  The freeze-out of molecules from the gas to the solid phase may play a role in defining the location and intensity of emission from different molecules \citep{Meijerink2008pL57}, and the settling of dust to the midplane of the disk can lead to drastic differences in the gas-to-dust ratio in the surface regions of disks at different evolutionary stages. We first tested a density-dependent freeze-out scenario (Model 2), where the molecules are assumed to freeze out on dust grains at a specific temperature which depends on the density of the gas, as described by \citet{Pontoppidan2006pL47}. The freeze-out usually starts at $\sim$1 AU, depending on which type of disk is modeled, and the boundary follows the disk horizontally out to large radii. Additionally, we tested an enhanced freeze-out model known as the ``vertical cold-finger effect" (\citet{Meijerink2009p1471}; Model 3) which models the diffusion of gas to below a horizontal freeze-out boundary by setting the freeze-out location based on the vertical disk temperature structure at a specific radius (see Figure 10 in that paper). In Model 4 we decreased the gas-to-dust ratio by an order of magnitude from 12800 to 1280.

The final model we examine is a disk + disk wind model, which is based on the results presented by \citet{pontoppidan2011p84} showing that a Keplerian disk + disk wind can better describe both the line profiles of the broad-based single-peaked CO emission lines observed in a number of bright T Tauri stars, as well as their asymmetric spectro-astrometry signal. The disk wind is a wide-angle non-collimated wind flowing over the surface of the disk that adds a sub-Keplerian velocity distribution to the standard Keplerian disk profile, which mainly contributes an approximately Gaussian core in the inner region of the line profile.  The disk wind model consists of gas flowing off of the disk along a set of streamlines with a locus point below the disk at a distance $d$ from the centre of the star (see Figure 10 in \citet{pontoppidan2011p84}); this distance $d$ as well as the other specific parameters for the structure of the disk wind are taken from values derived by \citet{pontoppidan2011p84} to reproduce the 4.7\,$\mu$m CO emission line profiles for AS 205 A (for a detailed description of the disk + disk wind model see \citet{pontoppidan2011p84} and \citet{Kurosawa2006p580}).  We keep two parameters, the mass loss rate from the disk and effective temperature for the central star (which sets the irradiating flux onto the disk surface), as free parameters in order to allow us to reproduce the central peak of the observed molecular line shapes properly (see the discussion of the modeling results below for more details).

Several additional components for accurately calculating molecular abundances and temperatures are not yet included in our modeling. The disk gas is currently assumed to be coupled with the dust, i.e. the gas temperature is set to be the same as the dust temperature; however, the gas and dust temperatures may be significantly different, although this is difficult to model properly (e.g., \citealt{Woitke2009p383, Gorti2008p287, Glassgold2009p142}). Also, to calculate the emission intensity for each transition of a specific molecule, the level populations must first be calculated assuming a specific excitation mechanism.  The standard formulation assumes local thermodynamic equilibrium (LTE), in which all the level populations are defined by the local temperature at that grid point.  However, non-LTE excitation processes such as radiative excitation due to the central star have been shown to be important for OH \citep{mandell2008pL25} whereas collisional excitation controls the H$_2$O rotational level populations \citep{Meijerink2009p1471}. Such processes will be described for HCN in Bast et al. 2011 (in prep). However, a non-LTE formulation for multiple molecules is time-consuming and difficult to implement. Moreover, collisional rate coefficients for vibration-rotation transitions are largely unknown. We therefore use only LTE excitation in our modeling and reserve the implementation of non-LTE excitation for a future analysis. Since our model is incomplete, we leave an exhaustive parameter study for future work; however, we can begin to compare line intensities and profiles with the data and derive preliminary molecular mixing ratios based on a realistic disk structure.

\subsection{Disk Modeling Results}
\label{result_model}

\begin{figure}[htb]
\centering
{
 \includegraphics[width=170mm]{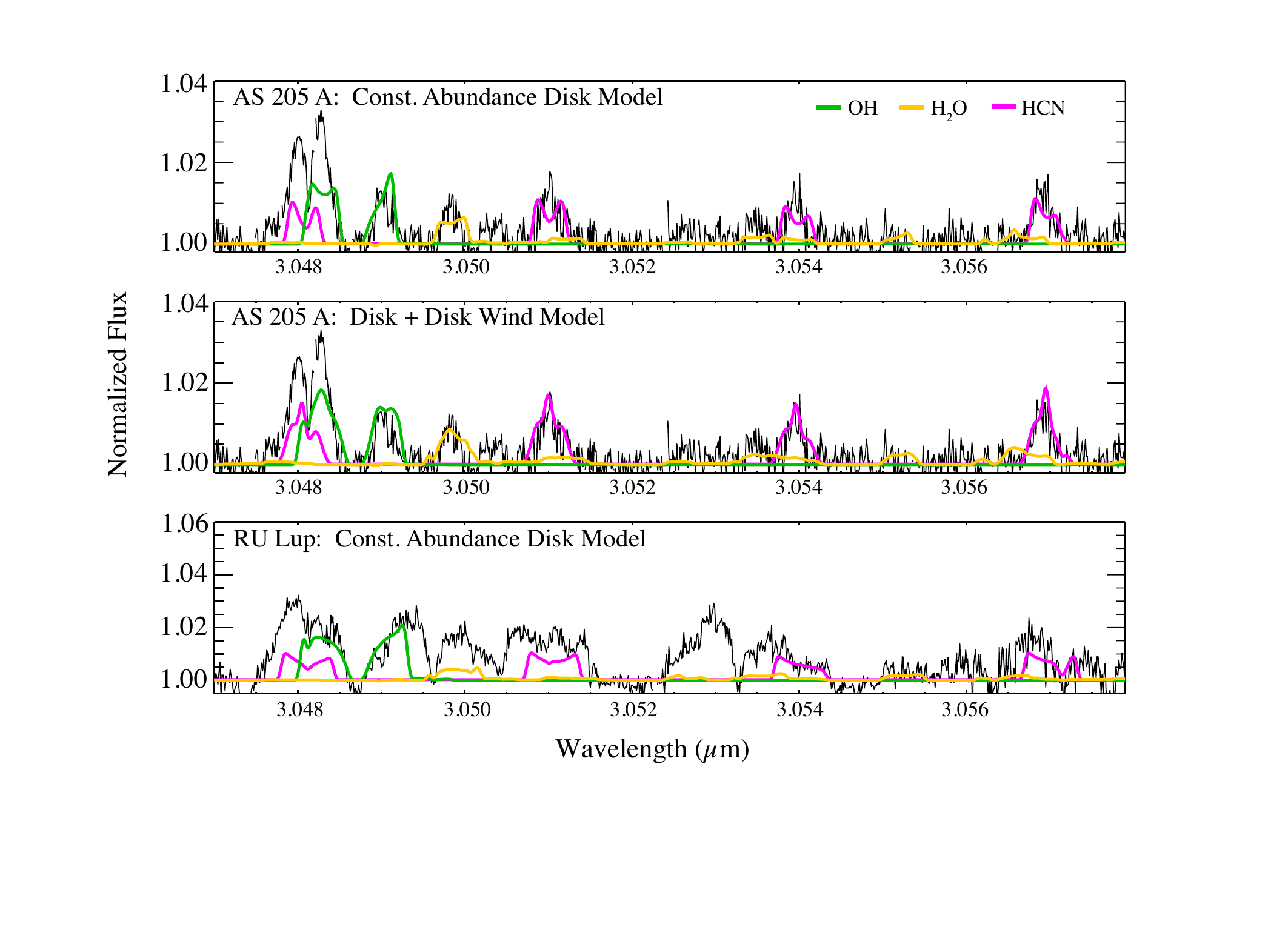}
 \caption{RADLite disk model results overplotted on the CRIRES
   data of AS 205 A (top and middle panels) and RU Lup (bottom
   planel). The top and bottom panels show results for the disk model only; the
   middle shows the disk + disk wind model. The disk + disk wind model
   clearly fits the centrally peaked line shape for AS 205 A much better than the disk-only model. }
 \label{fig:rulup_as205}
}
\end{figure}

\begin{figure}[htb]
\centering
{
 \includegraphics[width=170mm]{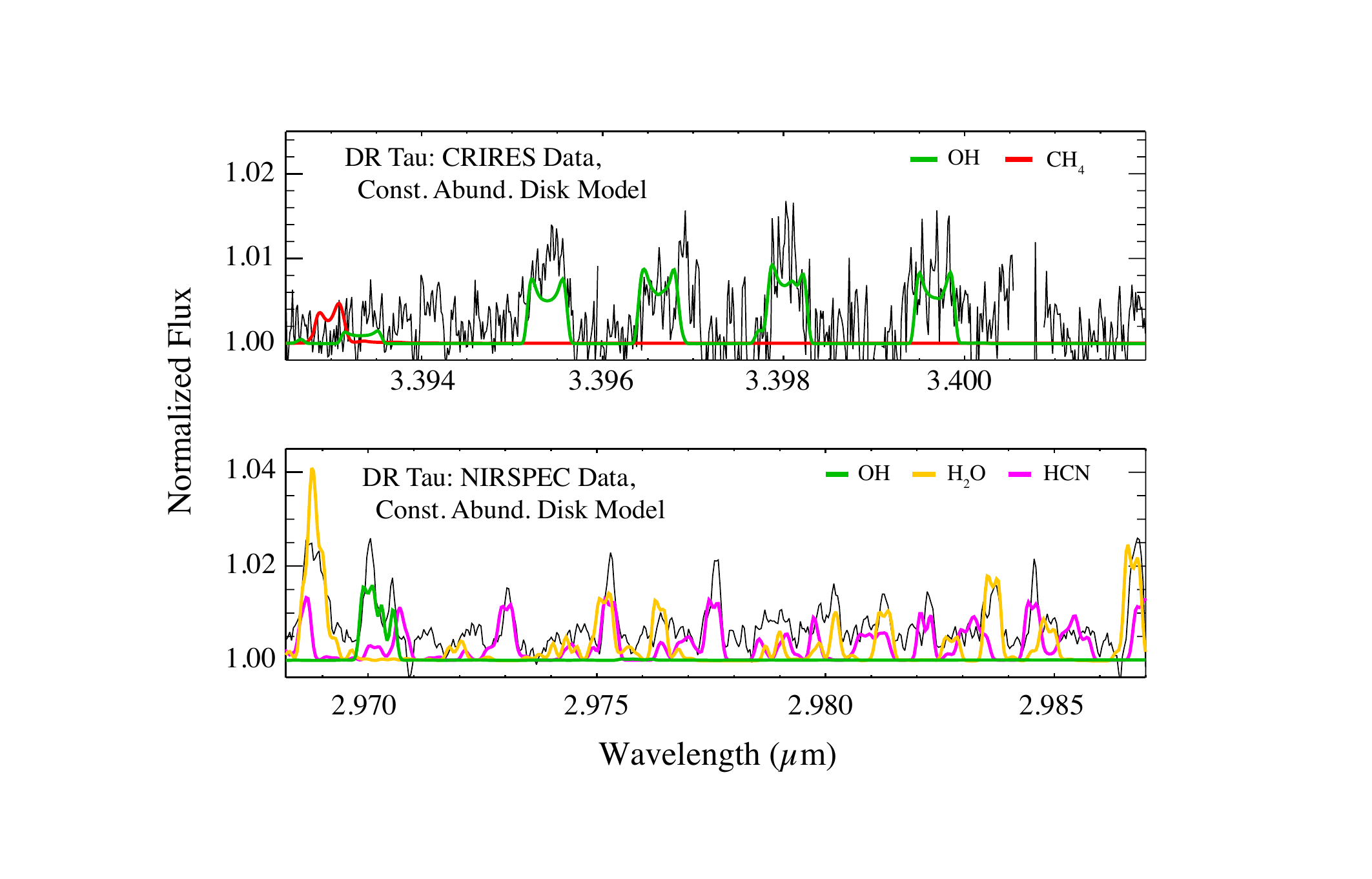}
 \caption{RADLite disk model results overplotted on the CRIRES
   (top) and NIRSPEC (bottom) data for DR Tau. Note that at the
   NIRSPEC resolution the discrepancy between model and observed line shapes is less noticeable.}
 \label{fig:drtau_col}
}
\end{figure}

\begin{figure}[htb]
\centering
{
 \includegraphics[width=110mm]{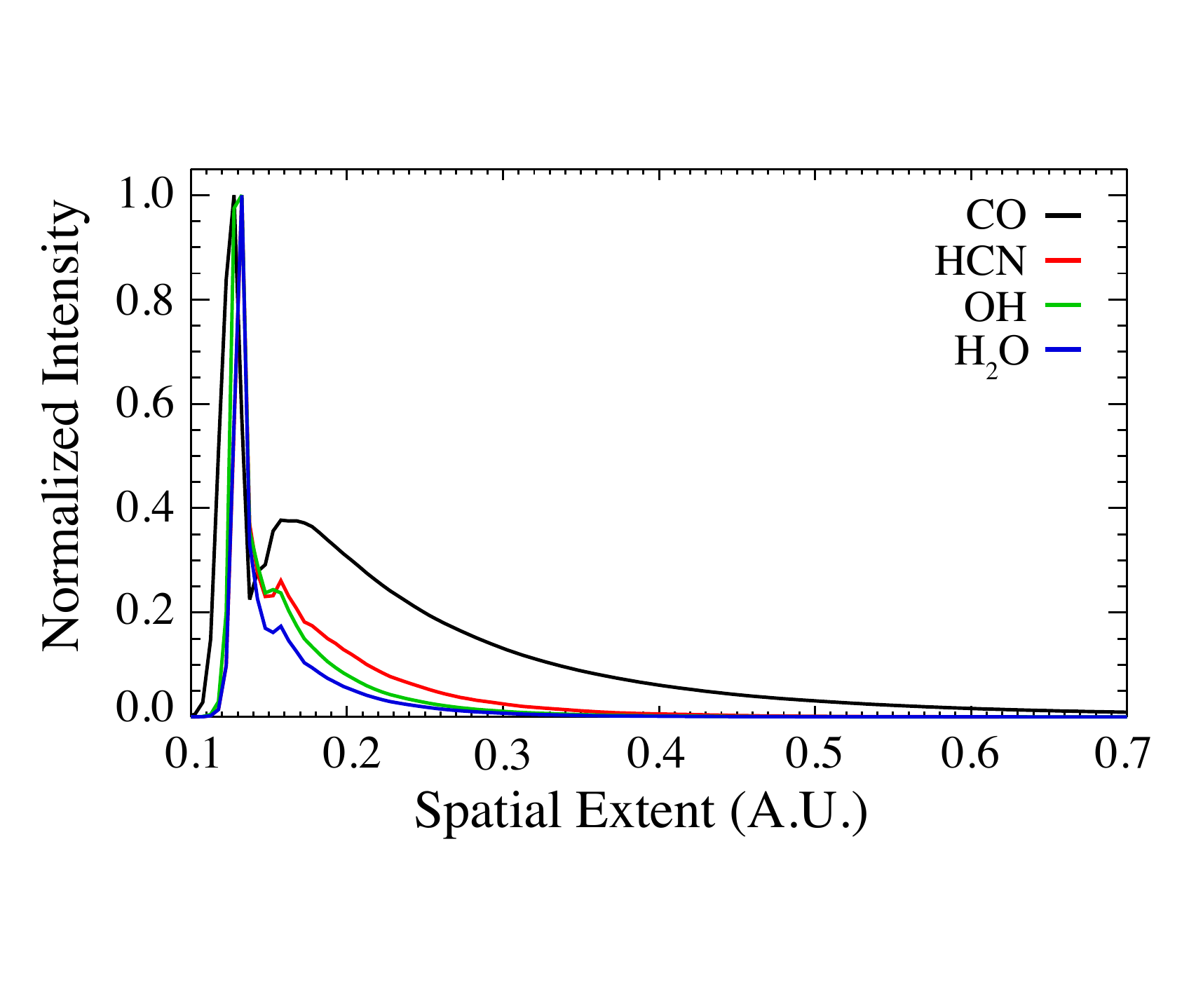}
 \caption{Distributions of the four primary molecules derived from disk modeling with RADLite. All of the 
 molecules emit primarily from within 0.2 AU, and only CO has emission extending out to 0.5 AU.}
 \label{fig:moldistrib}
}
\end{figure}

The top and bottom panels in Figure~\ref{fig:rulup_as205} present the best fit models for AS 205 A and RU Lup for HCN, H$_2$O and OH using Model 1 for a constant abundance distribution; Figure~\ref{fig:drtau_col} presents HCN and OH models for DR Tau.  The fiducial model produces a double-peaked line profile for all of the molecular species since the bulk of the line intensity originates from the hot inner disk (see Figure~\ref{fig:moldistrib}). Since the double-peaked model line profile clearly cannot fit the observed single-peaked line shapes, molecular abundances using the disk-only model are therefore obtained by matching the integrated intensity of the model lines with those of the observed lines; relative abundance ratios (compared with H$_2$O) are presented in Table \ref{tab:ratios}. The best fit for the constant abundance model for AS 205 A resulted in abundances (with respect to H$_2$) of $x_{\rm{HCN}}$ = 2$\times10^{-7}$, $x_{\rm{OH}}$ = 6$\times10^{-7}$ and $x_{\rm{H_2O}}$ = 3$\times10^{-6}$; however, due to uncertainties in the emission mechanism and the sensitivity of absolute abundances to specific model parameters, abundance ratios provide a more reliable measure for comparison between different objects.  Our model as presented in Figure~\ref{fig:rulup_as205} does not include higher vibrational excitation lines ($v'>1$) for HCN from the line list - the LTE model produces detectable lines for excited vibrational transitions but these are not observed for any of these sources. This is likely due to non-LTE excitation of the emitting gas, which produces sub-thermal intensities for higher-energy transitions as shown, for example, for water in \citet{Meijerink2009p1471}.

The CO abundance was once again derived by fitting the $^{13}$CO lines from our previous CRIRES observations and then multiplying the inferred abundance by 65; for AS 205 A we find $x_{\rm{CO}}$ = 1.95$\times10^{-5}$. For CH$_4$, an upper limit was obtained for DR Tau, but upper limits were not calculated for NH$_3$ and C$_2$H$_2$ due to the complexity of the spectral region where these molecules are present and the uncertainties in factors such as the line intensities of components within the NH$_3$ Q branch. Considering the similarity in the abundance ratios for the detected molecules using the slab model compared with the full disk model, we determined that any upper limits calculated for these molecules using the full disk model did not add useful information and might produce spurious results.

\begin{figure}[htb]
\centering
{
 \includegraphics[width=170mm]{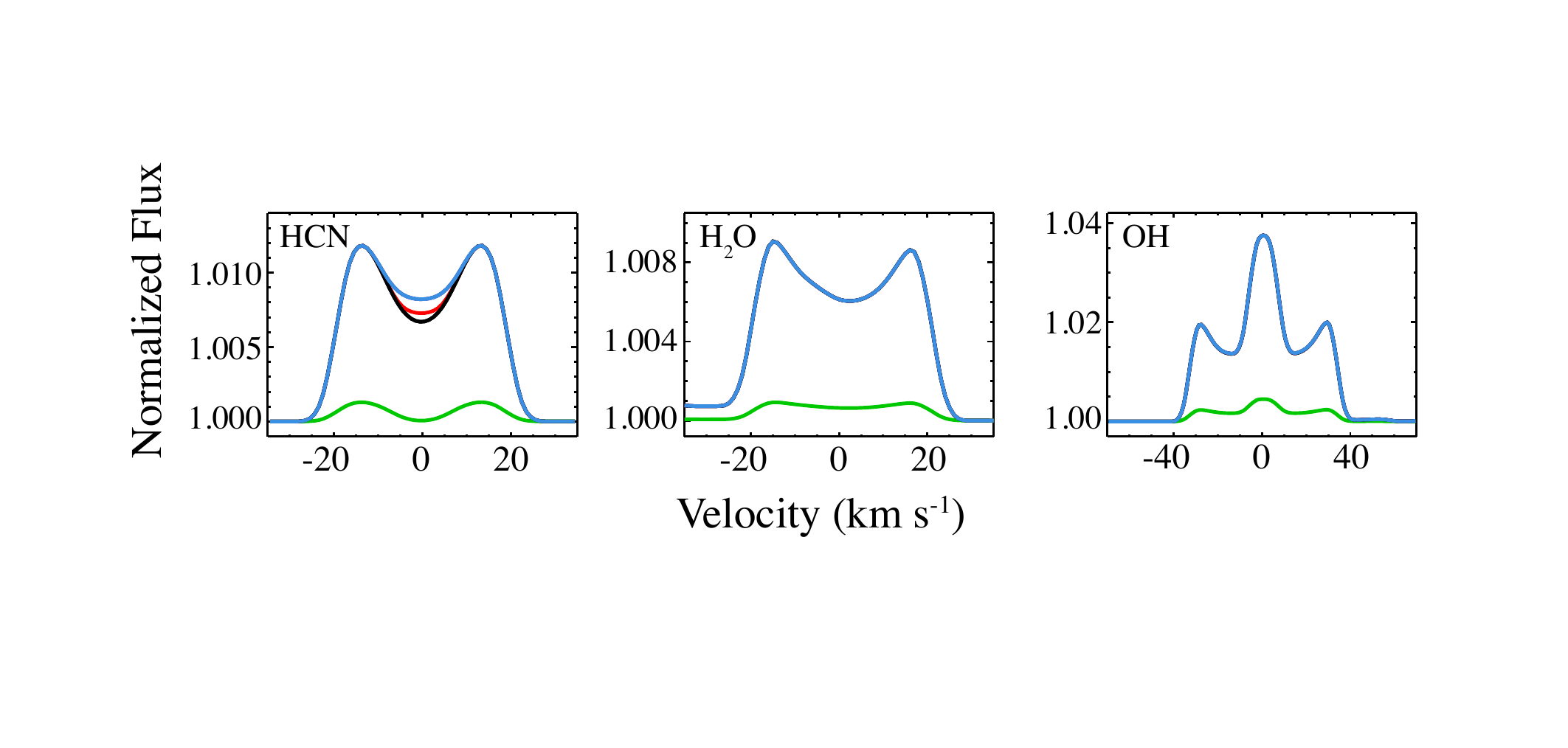}
 \caption{Radiative transfer disk models of HCN, OH and H$_2$O including parameters for AS 205 A. Black = constant abundance (Model 1), red = freeze out (Model 2), blue = ``vertical cold finger" model (Model 3) and green = gas-to-dust ratio of 1280 (Model 4; all other models using gas/dust = 12800). The HCN line shows a slight variation for Models 2 and 3 in the line core due to freeze-out at large distances, but emission from the other molecules originates from within the freeze-out distance (the red and black lines underlie the blue line).  The emission for the low gas-to-dust model is decreased in amplitude but the line shape does not change.}
 \label{fig:spec_models}
}
\end{figure}

As mentioned previously, we tested three disk models in addition to our fiducial constant abundance disk model: temperature- and density-dependent freeze-out (Model 2), a vertical cold finger effect (Model 3), and a lowered gas-to-dust ratio (Model 4). Fig~\ref{fig:spec_models} demonstrates that the line shapes for Models $1-3$ are very similar, due to the fact that the detected line emission in this wavelength region originates from within a few AU and is not modified by either the vertical cold finger effect or the freeze-out (i.e. the molecular condensation fronts), which only affect emission from the outer disk. The HCN line shows a slight decrease in depth for the cold finger model due to the reduced self-absorption of the colder gas; this effect is not seen for the H$_2$O and OH since the near-IR emission originates from gas within a radius which is smaller than the condensation radius for these molecules even when using the constant abundance model (this is in contrast with the mid-IR lines modeled by \citet{Meijerink2009p1471}). Lowering the gas-to-dust ratio from 12800 to 1280 (Model 4) clearly decreases the line intensity due to the decrease in the amount of emitting gas present above the optically thick dust layer, therefore requiring higher molecular abundances to fit the data. The final abundance ratios between H$_2$O and OH, HCN, C$_2$H$_2$ are not affected by the lowered gas-to-dust ratio but the CO abundances is more sensitive, most likely due to the higher abundance of CO across a large vertical scale height (i.e. \citet{Walsh2010p1607}).

\begin{figure}[htb]
\centering
{
 \includegraphics[width=110mm]{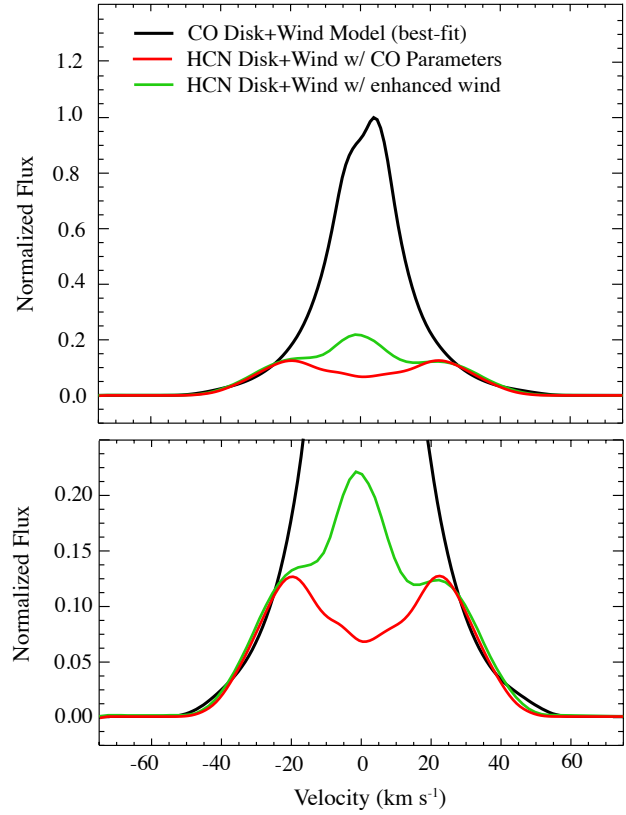}
 \caption{A CO (black) and HCN (red) line modeled using the disk wind
   model parameters for AS 205 A found by \citet{pontoppidan2011p84}. The HCN line should show a double-peaked profile, but our data clearly shows a profile very similar to the CO line profile. To achieve a centrally-peaked profile (green), we require either a higher mass-loss rate or increased stellar irradiation compared with the values derived for the CO lines.}
 \label{fig:diskwind}
}
\end{figure}

The middle panel in Figure~\ref{fig:rulup_as205} show the disk + disk wind model for AS 205 A, as discussed above. We originally calculated a model using the parameters for the disk wind derived by \citet{pontoppidan2011p84} for these sources from spectro-astrometry of the strong CO features at 4.7\,$\mu$m.  We expected this combined model to produce the most accurate match to the single-peaked line profiles seen in the data.  Interestingly, while the fit was better than the disk-only models, this model still produces a double-peaked line profile for HCN and H$_2$O, while the CO line profile has a central single peak, as illustrated in Figure~\ref{fig:diskwind}. This most likely indicates that the temperature in the disk wind model derived from the CO lines is too low to excite HCN and H$_2$O sufficiently. We were able to produce an adequate fit to the HCN and H$_2$O features using a wind model with either an enhanced mass loss rate ($\times3$) or an increased effective temperature for the central star ($\times2$); however, these solutions are clearly ad-hoc and only further emphasize the need for an improved treatment of the disk temperature structure (taking into account the different types of cooling and heating mechanisms) and the chemistry and infrared pumping of the wind. In addition to modeling the line shapes, spectro-astrometry of the HCN, OH and H$_2$O lines would help to further constrain the location and temperature of the emitting gas.

The molecular abundance ratios of $\sim$0.05 for HCN/H$_2$O and C$_{2}$H$_{2}$/H$_2$O and $0.2-0.3$ for OH/H$_2$O extracted from the models show little difference between the 3 sources, suggesting a common origin for the molecular material and similar excitation conditions.  This is not surprising, since we have chosen these three objects due to their similarly bright CO emission and centrally peaked line shape.  The relatively good agreement between the slab and disk models reinforces the conclusion by \citet{Meijerink2009p1471} that a slab model can produce reliable results because the emission arises from only a very narrow range of radii (and thus limited range of temperatures) in the disk model and because the emission is largely optically thin (the one exception is the OH line in RU Lup, which shows a complex line shape and therefore leads to somewhat divergent results between the two models). The most dissimilar abundance ratio between the different objects is CO/H$_2$O, which varies by a factor of 3 between the different sources. The fact that this ratio is larger than unity indicates that H$_2$O is not the major oxygen reservoir in the surface regions of the inner disk, assuming solar abundance ratios, but the derived values for the CO/H$_2$O ratio are dependent on the detailed temperature and density structure of the gas and the SED of the underlying dust continuum, and a more comprehensive modeling strategy utilizing a consistent model over a wide range of transition energies is needed to improve these constraints.

\section{Comparison with Chemical Models}
\label{discussion}

A number of models of inner disk chemistry exist in the literature, each with different levels of sophistication and each one taking different physical and chemical processes into account \citep{Markwick2002p632,   Agundez2008p831, Willacy2009p479, Glassgold2009p142, Walsh2010p1607, Najita2011p147}.  The models show a range of molecular ratios; typical HCN column densities relative to H$_2$O vary from 10$^{-4}$ - 10$^{-1}$ for the warm (T $>200$ K) molecular atmosphere of the disk, which agrees well with our results, while OH values relative to H$_2$O vary between 10$^{-5}$ - 10$^{-2}$, which is more than an order of magnitude below what we find. \citet{Bethell2009p1675} and \citet{Najita2011p147} have shown that the OH abundance is sensitive to photodissociation of parent molecules by both UV and X-rays, but detailed models including a complete treatment of the radiative environment have not been developed.  Additionally, as mentioned above, abundances from both our slab model and disk model are calculated using the assumption that the emission originates from an environment that can be described by LTE, but it has been proven previously by both observations \citep{mandell2008pL25,   Pontoppidan2010p887} and by modeling \citep{Meijerink2009p1471} that H$_2$O and OH rovibrational line emission in particular are almost certainly enhanced by radiative pumping from the central star, therefore making direct comparisons with models premature. The relative importance of LTE versus non-LTE excitation and de-excitation for all the relevant molecules requires detailed modeling and availability of collisional rate coefficients, and we leave these calculations to future work.

\citet{Najita2011p147} also explore the effects of larger dust grain sizes, lower accretion-related heating in the disk interior, and differing C/O ratios on the relative molecular abundances.  In particular, the ratios for carbon-bearing molecules are especially sensitive to the C/O ratio - values for the HCN/H$_2$O ratio vary dramatically from O/C = 0.25 ($5\times10^{5}$) to O/C = 2.0 (1$\times10^{-3}$). Our abundance ratios most closely match their calculated values for O/C = 1.5, with O/C = 1 producing values that are clearly larger than what we find.  C$_2$H$_2$ abundances vary across an even wider range, and additional processes such as the destruction of PAHs at high temperatures \citep{Kress2010p44} may also affect the abundance of acetylene.  The detailed impact of processes such as grain destruction and transport on the local O/C ratio and molecular abundances is still largely unexplored, and we leave any additional analysis to future work.

Our reported upper limits for CH$_4$ and NH$_3$ provide complementary information to the abundance ratios derived for the simpler organic species. If these molecules are formed primarily through grain surface chemistry in cold gas and then transported to warmer regions, they should then exhibit a different radial and vertical profile compared with HCN and C$_2$H$_2$.  Our observed limit on CH$_4$/H$_2$O of $<0.07-0.09$ is close to the average CH$_4$/H$_2$O ice abundance ratio of 0.05 found toward low-mass protostars, whereas our NH$_3$/H$_2$O limits are a factor of $3-4$ larger \citep{Oberg2011p98}. Our limits for CH$_4$/CO for DR Tau are also similar to upper limits found by \citet{Gibb2004pL113} and \citet{Gibb2007p1572} in an edge-on disk, suggesting that thermal desorption of CH$_4$ from grains is not efficient in the surface regions of the disk or that CH$_4$ has been transformed into other species.  However, both C$_2$H$_2$, HCN and CH$_4$ can also be produced effectively by high-temperature gas-phase reactions above $\sim 600$~K \citep{Doty2002p446}. \citet{Agundez2008p831} predict large CH$_4$ and NH$_3$ abundances very close to the star (R $<0.5$ AU), with a sharp drop-off as the gas temperature decreases.  Their model ratios with respect to HCN or H$_2$O are consistent with our observed upper limits except for the innermost rim at $\sim$0.1 AU where the model CH$_4$ abundance becomes higher than that of HCN.  \citet{Willacy2009p479} predict similarly low abundance ratios of both CH$_4$ and NH$_3$, and attribute the low NH$_3$ values to efficient conversion to HCN in the warm inner disk.

In future work we will expand our modeling with RADLite to explore additional parameter space with regard to these disk characteristics. We will also augment our radiative transfer modeling to include the contribution from non-LTE radiative pumping, in order to allow us to model both the NIR and MIR emission in a self-consistent manner. Another important constraint when modeling emission lines is the gas temperature, which here is set to be the same as the dust temperature. Both \citet{Glassgold2001p251} and \citet{Kamp2004p991} show that the gas is decoupled from the dust, hence future models that can include different temperatures for the dust and the gas would be better able to model the molecular line emission from T Tauri stars.  

\section{Conclusion}
\label{concl} 
We have successfully detected emission from organic molecular gas tracers in T Tauri disks in the NIR for the first time. Emission from HCN as well as H$_2$O and OH was detected from all three sources, and we detected emission from several transitions of C$_2$H$_2$ in AS 205 A. Strong upper limits were placed on emission from CH$_4$ in DR Tau and on NH$_3$ in AS 205 A and RU Lup.  Based on models, the emission likely arises from the inner $0.1-1$ AU of the disk. These first detections demonstrate the feasibility of constraining molecular abundance ratios and temperatures for organic molecules in the planet-forming regions of circumstellar disks using spectrally resolved NIR transitions, and allow us to begin to constrain models of disk chemistry, dynamics and structure. These results provide a direct comparison to observations of unresolved emission from the same molecules observed in the same sources with the {\it Spitzer} telescope and upcoming observations with the Atacama Large Millimeter Array (ALMA).

 \begin{acknowledgements}
 AMM is supported by the Goddard Center for Astrobiology. JEB is supported by grant 614.000.605 from Netherlands Organization of Scientific Research (NWO). EvD acknowledges support from a NWO Spinoza Grant and from Netherlands Research School for Astronomy (NOVA). The authors are very grateful to Klaus Pontoppidan for his central role in the CRIRES observations and for making his RADLite program available. They also thank Daniel Harsono for help with the disk modeling, and the anonymous referee for helpful suggestions for improving the manuscript. This research made use of the ESO/ST-ECF Science Archive Facility. Some of the results herein were obtained at the W. M. Keck Observatory, which is operated as a scientific partnership among the California Institute of Technology, the University of California, and NASA. The Observatory was made possible by the generous financial support of the W. M. Keck Foundation.
 \end{acknowledgements}


\clearpage

  \begin{table*}
\footnotesize
\caption{Summary of Observations}
 \begin{minipage}[t]{\columnwidth}
 \renewcommand{\footnoterule}{}  
 \centering
  \thispagestyle{empty}
  \begin{tabular}{l c c c c c c}
  \hline
   \hline
   Star & Dates & Instrument & Wavelength Range & t$_{int}$  & Standard
   & Molecules \\
        &  & & ($\mu$m) & (min) & \\
   \hline  
AS 205 A & UT July 23 2009              & VLT/CRIRES     & $2.995 -  3.058$ & 36 & BS6378 & HCN, C$_2$H$_2$, NH$_3$, \\
   & & & & & & H$_2$O, OH \\
    \vspace{1 pt}\\
DR Tau & UT October 30 2007      & Keck/NIRSPEC & $2.947 - 2.987$ & 28  & HR1620& HCN, H$_2$O, OH \\      
               & UT December 31 2008 & VLT/CRIRES      & $3.333 - 3.414$ & 48 & HR2421 & CH$_4$, OH \\      
    \vspace{1 pt}\\
RU Lup & UT August 5 2009         & VLT/CRIRES       & $2.995 - 3.071$ & 32 & BS5812 &  HCN, C$_2$H$_2$, NH$_3$, \\
   & & & & & & H$_2$O, OH \\  
\hline
\end{tabular} 
 \end{minipage}
\label{tab:obs}
\end{table*}

  \begin{table*}
\footnotesize
\caption{Line Intensities for Selected Molecular Features}
 \begin{minipage}[t]{\columnwidth}
 \renewcommand{\footnoterule}{}  
 \centering
  \thispagestyle{empty}
\begin{tabular}{l l c c c}
  \hline
   \hline
   Source & Molecule & $\lambda_{rest} (\mu$m) & Transition\footnote{Multiple transitions are listed if a single molecular feature is an unresolved blend or a doublet} & Intensity (W/cm$^2$)\\
   \hline  
   AS 205 A  & HCN & 3.00155   & ($\nu_1+\nu_2+\nu_3$)-($\nu_2+\nu_3$), R13 e &  2.1$\times10^{-22}$ \\
    & & 3.00158 & $\nu_1$, R6 & 4.7$\times10^{-22}$ \\
    & H$_2$O & 3.02066 & (100)6$_{51}$ - (000)7$_{62}$ & 1.4$\times10^{-22}$ \\
    & & 3.02067 & (100)6$_{51}$ - (000)7$_{61}$ & 4.3$\times10^{-22}$ \\
    & OH & 2.99995 & X$^{2}\Pi_{1/2}$ (1-0), P5.5 f & 1.2$\times10^{-21}$ \\
    & & 3.00022 & X$^{2}\Pi_{1/2}$ (1-0), P5.5 e & 1.2$\times10^{-21}$ \\
    & C$_2$H$_2$ & 2.99800 & $\nu_{2}+\nu_{4}+\nu_{5}$, R23e & 1.4$\times10^{-22}$ \\
    & & 2.99801 & $\nu_3$, R17 e & 1.4$\times10^{-22}$ \\
    \vspace{1 pt}\\
   RU Lup & HCN & 3.00155 & ($\nu_1+\nu_2+\nu_3$)-($\nu_2+\nu_3$), R13 e & 1.5$\times10^{-22}$ \\
    & & 3.00158 & $\nu_1$, R6 & 3.4$\times10^{-22}$ \\
    & H$_2$O & 3.02066 & (100)6$_{51}$ - (000)7$_{62}$ & 6.7$\times10^{-23}$ \\
    & & 3.02067 & (100)6$_{51}$ - (000)7$_{61}$ & 2.1$\times10^{-22}$ \\
    & OH & 2.99995 & X$^{2}\Pi_{1/2}$ (1-0), P5.5 f & 2.5$\times10^{-21}$ \\
    & & 3.00022 & X$^{2}\Pi_{1/2}$ (1-0), P5.5 e & 2.5$\times10^{-21}$ \\
    & C$_2$H$_2$ & 2.99800 & $\nu_{2}+\nu_{4}+\nu_{5}$, R23 e & 1.5$\times10^{-22}$ \\
    & & 2.99801 & $\nu_3$, R17 e & 1.3$\times10^{-22}$ \\
    \vspace{1 pt}\\
  DR Tau  & HCN & 2.97525 & $\nu_1$, R17 & 1.4$\times10^{-22}$ \\
   & & 2.97528 & ($\nu_1+2\nu_2$)-($2\nu_2$), R 33 & 4.1$\times10^{-24}$ \\
   &H$_2$O & 2.95642 & (001)12$_{66}$ - (000)13$_{67}$ & 1.4$\times10^{-22}$ \\
   & OH & 2.96043 & X$^{2}\Pi_{1/2}$ (1-0), P4.5 e & 3.1$\times10^{-22}$ \\
   & & 2.96026 & X$^{2}\Pi_{1/2}$ (1-0), P4.5 f & 3.1$\times10^{-22}$ \\
   & CH$_4$ & 3.38053 & $\nu_3$, P6 (F2-F1) & $<6.0\times10^{-23}$ \\
   & & 3.38040 & $\nu_3$, P6 (F1-F2) & $<6.0\times10^{-23}$ \\
\hline
\end{tabular} 
 \end{minipage}
\label{tab:lines}
\end{table*}

  \begin{table*}
\footnotesize
\caption{Molecular ratios, relative to H$_2$O, estimated by using the slab model and constant 
abundance disk model (Model 1), respectively. Abundances for DR Tau (except for CH$_4$) were calculated from
data obtained with the NIRSPEC instrument on Keck; all other abundances were determined from data obtained with the CRIRES instrument on the VLT.}
 \begin{minipage}[t]{\columnwidth}
 \renewcommand{\footnoterule}{}  
 \centering
  \thispagestyle{empty}
\begin{tabular}{l l c c c c c c c}
  \hline
   \hline
   Model & Source & HCN/H$_2$O & OH/H$_2$O & CH$_4$/H$_2$O & NH$_3$/H$_2$O & C$_2$H$_2$/H$_2$O  & CO/H$_2$O & FWHM \\
    & & & & & & & & (km s$^{-1}$) \\
   \hline  
Slab$^a$& AS 205 A & 0.06$\pm$ 0.01 & 0.25$\pm$ 0.03 & - & $< 0.16$ & 0.05$\pm$ 0.01 & 11.6 & 22.5\\   
          & DR Tau & 0.05$\pm$ 0.01 & 0.16$\pm$ 0.02 & $< 0.07$ & - & $<0.03$ & 29.0 & 10.8\\      
         & RU Lup & 0.05$\pm$ 0.02 & 0.3$-$0.6$^b$ & - & $<0.2$ & 0.05$\pm$ 0.02 & 18.3 & 16.2\\  
    \vspace{1 pt}\\
Disk & AS 205 & 0.07 & 0.2 & - & - & - & 6.5 \\
        & DR Tau & 0.07 &  0.13 & $< 0.09$ & - & - & 2.9 \\
        & RU Lup & 0.08 & 1.5 & - & - & - &  3.3 \\
\hline
\footnotetext[1]{Slab model abundances were calculated with a common temperature of 900 K, except for CO which was best fit with a temperature of 600 K}
\footnotetext[2]{The OH abundance ratio from the slab model for RU Lup is given as a range, based on either a Gaussian or Keplerian fit.}
\end{tabular} 
 \end{minipage}
\label{tab:ratios}
\end{table*}

  \begin{table*}
\footnotesize
\caption{Parameters used in each specific fiducial disk model.}
 \begin{minipage}[t]{\columnwidth}
 \renewcommand{\footnoterule}{}  
 \centering
  \thispagestyle{empty}
  \begin{tabular}{l l l l}
  \hline
  \hline
  Parameters\footnote{References: (1) \citet{Andrews2009p1502} ; (2) \citet{Ricci2010p15}; (3) \citet{Andrews2007p705}; (4) \citet{Schegerer2009p367}; (5) \citet{Robitaille2007p328}; (6) \citet{Isella2009p260}, (7) \citet{pontoppidan2011p84}, (8) \citet{Herczeg2008p594}, (9) \citet{Lommen2007p211}, (10) set to be the same as for AS 205 A however this parameter do not impact strongly on the line intensities, (11) Increased from 0.1 AU as found by \citet{Schegerer2009p367} in order to fit the narrow line shapes with the published inclination} & & Source & \\
   \hline
   & AS 205 A & DR Tau & RU Lup \\
   $M_\star$ [$M_\odot$] & 1.0$^1$ & 0.8$^2$ & 0.7$^7$ \\
   $R_\star$ [$R_\odot$]  & 3.7$^1$ & 2.1$^2$ & 1.3$^8$ \\
   $T_{eff}$ [K] & 4250$^1$  & 4060$^2$ & 4060$^8$ \\
   $M_{disk}$ [$M_\odot$] & 0.029$^1$ & 0.01$^3$ & 0.03$^9$ \\
   $R_{out}$ [AU] & 200$^1$ & 100$^3$ & 100$^4$ \\
  $R_i$ [AU] & 0.14$^1$ & 0.05$^4$ & 0.2$^{11}$ \\	
   $h_{p}$/$R$ & 0.21$^1$ & 0.15$^8$ &  0.20$^8$ \\
    $\alpha$ & 0.11$^1$ & 0.07$^8$ & 0.09$^8$ \\
    $\beta$ & -0.9$^1$ & -0.5$^3$ & -0.9$^{10}$ \\
    $i$ [$\degr$] & 20$^7$ & 9$^7$ & 35$^7$ \\   
    \hline
\end{tabular} 
 \end{minipage}
\label{tab:parameters1}
\end{table*}

\end{document}